\documentclass[pra,showpacs,groupedaddress,superscriptaddress,twocolumn]{revtex4}

\usepackage{graphicx}
\usepackage{amsmath}
\usepackage{amsfonts}
\usepackage{amssymb}
\usepackage{bbm}
\usepackage{times}
\def\be{\begin{equation}}
\def\ee{\end{equation}}
\def\bea{\begin{eqnarray}}
\def\eea{\end{eqnarray}}
\newcommand{\ket}[1]{|#1\rangle}
\newcommand{\bra}[1]{\langle#1|}

\begin{document}

\title{Quantum Phase Estimation with Entangled Photons produced by Parametric Down Conversion}

\author{Hugo Cable}
\email{cqthvc@nus.edu.sg}
\affiliation{Centre for Quantum Technologies, National University of
Singapore, 3 Science Drive 2, Singapore 117543}

\author{Gabriel A.\ Durkin}
\email{gabriel.durkin@qubit.org}
\affiliation{Quantum Laboratory, NASA Ames Research Center,  Moffett Field, California 94035, USA}

\date{\today}

\begin{abstract}
We explore the advantages offered by twin light beams produced in parametric down-conversion for precision measurement. The symmetry of these bipartite quantum states, even under losses, suggests that monitoring correlations between the divergent beams permits a high-precision inference of any symmetry-breaking effect, e.g. fiber birefringence.  We show that the quantity of entanglement is not the key feature for such an instrument. In a lossless setting, scaling of precision at the ultimate `Heisenberg' limit is possible with photon counting alone. Even as photon losses approach 100\% the precision is shot-noise limited, and we identify the crossover point between quantum and classical precision as a function of detected flux.  The predicted hypersensitivity is demonstrated with a Bayesian simulation.
\end{abstract}


\maketitle
\section{introduction}
We wish to promote the fitness of two spin-$j$ systems combined in an overall spin zero singlet as a resource for metrology. A photonic implementation of such states is readily and scalably generated using stimulated parametric down-conversion (PDC) \cite{EntangledPhotonLaser}.  Both their rotational symmetry \cite{RotInv} and persistence of entanglement \cite{Eisenberg04,Durkin04} under photonic loss channels has made the singlets natural candidates for quantum key distribution \cite{Durkin02}. In this letter we further highlight their utility, this time in a parameter-estimation protocol. Generally, such protocols are rated by the precision (or uncertainty) associated with unbiased estimation of the unknown parameter, and how quickly this precision is lost under relevant decoherence. We will show that given ideal conditions the singlets allow a precision scaling at the Heisenberg limit (the ultimate limit for linear quantum processes, and for which noise scales as $1/N$ with respect to the light intensity or particle number $N$). Under incoherent photon loss measurement precision is naturally degraded, but at a much gentler rate than other proposals \cite{Chen07} where the decay can be exponential in $N$. (Recently, the role of photon losses in optical precision experiments  was examined carefully \cite{Kacprowicz09}.)

Consider an instrument  broken down into three components \cite{Durkin10}: pure probe state $\ket{j}$, unitary evolution $U\!=\! \exp(-i\phi\hat{H}t)$ under a time-independent hermitian Hamiltonian $\hat{H}$, and complete projective measurements, $\hat{M}\!=\!\sum_i m_i\ket{i}\bra{i}$, where $\langle i\vert j\rangle\!=\!\delta_{ij}$.  We wish to infer evolution time $t$ from frequencies of outcomes $m_i$, but
$t$ may equally represent an interferometer phase, a magnetic or gravitational field, or some other real-valued continuous
variable.  Extrapolating from a set of parameter-dependent measurements to an estimation of that parameter can be a challenging task. The probability distributions for individual outcomes are often non-Gaussian,  having multiple peaks or broad tails. In addition to designing a good estimator from measurement data, it is as important to employ measurements sensitive to small changes in the unknown parameter. An optimal measurement should also, if possible, achieve highest estimation precision for all parameter values. Conditions for identifying optimal measurements were identified in early work on quantum Fisher information. One parameter-independent approach uses canonical measurements, but these are hard to implement directly \cite{CanonicalMeasurements}. In other proposals, measurements are optimal near particular `sweet spots' in parameter space, and require multi-step adaptive measurements to exploit them \cite{Monras06}.

We first present a protocol in a decoherence-free setting for which our proposed measurements are optimal,
parameter-free and practicable; they may be realized in a laboratory by photon-counting or spin projections. We extend the
analysis to consider realistic decoherence.  Translating our protocols to a quantum optics setting, characterized by a PDC
source, we evaluate the effect of photon losses (in transmission and detection) on precision. It emerges that spin-projection
measurements are no longer optimal or parameter independent. Nonetheless, precision is always at least as good
as the classical upper bound (noise scaling $\propto N^{-1/2}$, called the shot-noise limit) for \emph{any} loss.

\section{Input States}  

Let us introduce the state in the spin representation \cite{notation}.  Our probe is a bipartite maximally-entangled spin singlet:
\begin{equation}\label{singlet}
| \psi_0^{(j)} \rangle\! =\! \frac{1}{\sqrt{2j+1}}\sum_{m= -j}^{+j} (-1)^{j-m} |j,m\rangle_{za} \otimes   |j,-m\rangle_{zb}
\end{equation}
% on the tensor product space $\mathbb{C}^{(2j+1)}_a \otimes \mathbb{C}^{(2j+1)}_b$,
labeling component spaces `$a$' and `$b$'. The state has total spin $J=0$: $\langle \psi_0^{(j)} | \vec{J}^2 | \psi_0^{(j)} \rangle = 0$ where $ \vec{J} = ( \vec{J}_a +  \vec{J}_b)$.   It is `rotationally' invariant \cite{RotInv} under $U^{(j)}_a(g_1) \otimes U^{(j)}_b(g_2)$ when $g_1=g_2 \in SU(2)$; its description in Eq.~\eqref{singlet} is identical in any spin basis, e.g. $z \mapsto x \mapsto y$, and it has properties that change only with the \emph{relative} transformation $\mathbbm{1}_a \otimes U^{\dagger (j)}_b (g_1) U^{(j)}_b(g_2)$. There is an application here for relative measurements made between non-local observers; global phenomena are excluded.

\begin{figure}\centering
\includegraphics[width=3.4in]{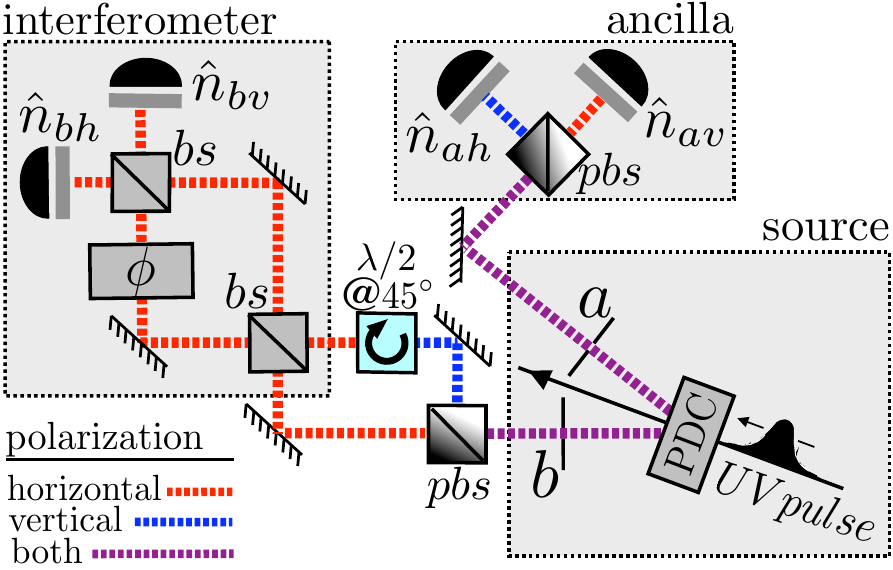}
\caption{\label{Apparatus}
Instrument composed of PDC source \cite{EntangledPhotonLaser},  linear-optical elements (mirrors,  phase element $\phi$, $50/50$ beam-splitters `$bs$', polarising beam-splitters `$pbs$', half-wave plate `$\lambda/2$'), and photocounters. In microscopy or interferometry, modes $a$ may be considered an ancilla/reference, only the photons in modes $b$ interact with the `sample'. The down-conversion Hamiltonian is $\hat{H}=\kappa(a_h^\dag b_v^\dag\!-\!a_v^\dag b_h^\dag)\!+\!h.c.\,$ with $\kappa$ a coupling strength for the nonlinear process, producing light in four optic modes via application to the vacuum: $\exp ( -i \hat{H} t ) | 0 \rangle$. Given effective interaction time $\tau\!=\!\kappa t$, and disregarding losses, the source produces a mixture of spin-$j$ singlets, Eq.\eqref{singlet}, with weighting $(2j\!+\!1)\tanh^{4j}\! \tau/\cosh^{4} \! \tau$, and each singlet has a photon number $N\!=\!4j$.  Overall, the intensity is $\langle \hat{N} \rangle\! =\!4\sinh^2\tau$ and $\Delta^2 \hat{N} \!=\! \cosh 4 \tau\!-\!1$.  Weightings of higher-photon-number singlets increase with $\tau$ but the large $\Delta^2 \hat{N}$ indicates a severe flattening of the distribution. We note the immunity of counting measurements to undesired path differences occurring outside of the `interferometer' block. %We suggest two simple improvements: firstly, the addition of a delay line and switch before the test sample in the $b$ modes allows the number of photons traversing the sample to be controllable,  `heralded' by the $a$ outcomes;  secondly, allowing both modes $a_h$ and $b_v$ to encounter the sample phase $\phi$, the effective Hamiltonian becomes $\hat{J}_{yb} \!-\! \hat{J}_{ya}$. This dual probe means a doubling of the signal to noise ratio; Fisher information for $\phi$ becomes $16j\left( j+1\right) /3$  (although twice as many photons now interact with the target.
For fiber calibration the apparatus would be simpler, comprising polarizing beam-splitters and photon detectors terminating both test  and reference fibers coupled to the PDC source  (without an explicit interferometer or $\lambda/2$ plate).}
\end{figure}

\section{Phase-Estimation Protocol}    

Examine now  the estimation of a relative phase $\phi$ accumulating between $a$ and $b$.  Breaking the symmetry of $| \psi_0^{(j)} \rangle$ by applying $\mathbbm{1} \otimes U^{(j)}_b$, yields the state $|\psi_\phi^{(j)} \rangle\!=\!\exp \left( -i \phi \hat{J}_{yb} \right) | \psi _{0}^{(j)} \rangle$ if we choose rotation about $y$. In the paradigm above, $\hat{H} \mapsto \hat{J}_{yb}$ and $t \mapsto \phi$. Then make projective measurements onto the basis of $z$-eigenstates for $a$ and $b$, i.e. $\hat{M} \mapsto \hat{J}_{za} \otimes \hat{J}_{zb}$.  The probabilities for the $\left(2j+1\right)^2$ measurement outcomes are
\begin{equation}
\label{ProbOutcome}
P_{AB}\left( \phi \right) =\left\vert \left\langle j,A\right\vert _{z,a}\left\langle j,B\right\vert _{z,b}\left\vert \psi_{\phi} \right\rangle \right\vert ^{2}\!=\!\frac{d_{B,-A}^{\left( j\right) }\left( \phi \right) ^{2}}{\left( 2j+1\right)}.
\end{equation}
Note that $P_{AB}\left( \phi \right)\!\!=\!\!P_{AB}\left( -\phi \right)$, and  to eliminate ambiguity the range of $\phi$ can be restricted to $[0,\pi]$.

\section{Quantifying Precision} 

To establish this scheme's performance we employ classical Fisher information $I_{cl}\left(\phi\right) \!=\!\sum_{A,B}P_{A,B}\left[ \frac{d}{d\phi }\ln \left( P_{A,B}\right) \right] ^{2}$.  It provides a distance metric in probability space, e.g for $\epsilon \ll 1$ the distance between $P_{AB}( \phi )$ and $P_{AB}( \phi +\epsilon)$ is $\epsilon \sqrt{I_{cl}}$. A lower bound on the variance of any unbiased estimator of $\phi$ is $I_{cl}^{-1}$ \cite{Durkin-Dowling}.  Given an unknown rotation by $\phi$ about the `$by$' axis  %the derivatives $dP_{AB}/d\phi$ are given by
%$\left[d^{\left( j\right)}_{B,-A}\left( \phi \right)/\left( 2j+1\right)\right]\times$
%$\left[ N^{(j)}_{-}\left(-A\right) d_{B,-A-1}^{\left( j\right) }\left( \phi \right) \!-\!N^{(j)}_{+}\left( -A\right) d_{B,-A+1}^{\left( j\right) }\left( \phi \right) \right]$.
 %summing over all detection outcomes
a short calculation (Appendix A) using Eq.\eqref{ProbOutcome} gives $I_{cl}\left( \phi \right) =4j\left( j+1\right)/ 3$, independent of $\phi$ and achieving the quadratic scaling characteristic of the Heisenberg limit (identifying a spin-$j$ state as a composite of $2j$ spin one-half particles, the singlet then has particle number $N \!=\! 4j$ and $I_{cl} \propto N^2$).

The quantum Fisher information $I_{qu}$, provides a saturable upper bound, $I_{qu} \geq I_{cl}$. It is a function only of probe and dynamics, assuming the best possible measurement without defining it explicitly.  For pure probe states,  $I_{qu} /4\!=\! \Delta^2 \hat{H} = \langle \hat{H}^2 \rangle \!-\!  \langle \hat{H} \rangle^2$  \cite{QFI}. Using again $\hat{H} \mapsto \hat{J}_{yb}$, $I_{qu}=4\langle \hat{J}_{yb}^{2}\rangle =4\langle \vec{J}_{b}^{2}\rangle /3=4j ( j\!+\!1) /3$ by symmetry.  Remarkably $I_{qu}\!=\!I_{cl}$, and our original measurement choice is optimal and independent of the value of the unknown parameter $\phi$, a preferred property of any parameter estimation scheme \cite{OptimalMeasurements,Durkin10}.
We make at this point some observations about entanglement. First, making a comparison between the probe $\vert \psi_0^{(j)} \rangle$ and  Bell states $\vert \psi_0^{(1/2)} \rangle$ numbering $2j$ (so total particle number $4j$ is the same), we see that the former has much less entanglement than the latter, $\log_2(2j+1)$ e-bits versus $2j$ e-bits, but a greater value for $I_{qu}$, $4j(j\!+\!1) /3$ versus $2j$.  Second, although the Hamiltonian only operates on the $b$ modes,  no phase information is imprinted locally as the reduced state $\rho^{(b)}$ is always maximally mixed -- nothing is learned by measurements exclusive to the space in which dynamics occur.  One might believe that the dependence of precision on bipartite measurements is due to entanglement. However, we will show it is retained under a disentangling decoherence channel -- exhibiting non-locality without entanglement.

\begin{figure}[!]\centering
\includegraphics[width=3.3in]{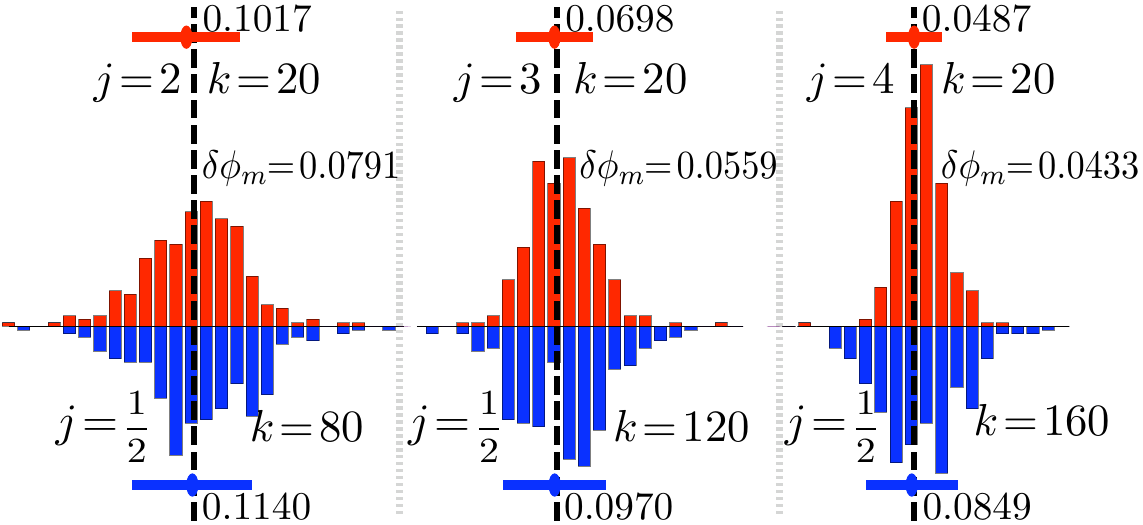}
\caption{\label{SIM} \small{Simulation with performance of spin-j singlets (red/above) juxtaposed with spin-$1/2$ Bell pairs (blue/below). Like coherent state light, the Bell states have a precision $ I_{qu}$ that scales linearly with particle number. There were $250$ independent phase estimates contributing data points to each of the six charts. A single estimate is obtained after a sequence of $k=20$ Bayesian updates for probes with $j\!=\!2,3,4$ using the distribution of Eq.\eqref{ProbOutcome}. The true value of $\phi$ is indicated by black dashed lines. Maintaining the same particle number resource for the Bell probes as the higher-$j$ singlets equates with keeping $j k $ constant.  Experimental error bars approach the theoretical limit $\delta \phi_m = 1/\sqrt{k I_{qu}}$ as particle number $1000 jk$ increases; red bars are smaller than corresponding blue ones (below them) indicating superior precision of the higher-spin singlets in this pre-asymptotic regime. }}
\end{figure}

\section{Simulation}

The quantum bound  $(\delta \phi)^2 \geq I_{qu}^{-1}$ on mean-squared error is attainable given infinite repetitions of the experiment (using maximum-likelihood estimation), but  superior convergence of the high-$j$ singlets may also be shown when the number of samplings is relatively small, the \emph{pre-asymptotic} regime. A  Bayesian protocol, considered in Fig.~\ref{SIM}, compares performance of higher-$j$ singlets  with Bell singlets, so that the same total energy is detected in both cases.  We use data sequences with $k$ updates of the (uniform) prior distribution $P_0=1/\pi$ given measurements $\{ A_{i},B_{i}\}= \{ A_{1},B_{1},  A_{2},B_{2},\dots ,A_{k},B_{k}\} $.  Each sequence leads to a single $\phi$ estimate that is the mean of the final posterior distribution $P_k$. The conditional probability $P(\{ A_{i},B_{i}\} |\phi )$ is determined from Eq.~(\ref{ProbOutcome}).  The update rule is $P_{k}\left( \phi | \{A_{i},B_{i}\} \right) \!\propto\! P( \{ A_{i},B_{i}\} |\phi ) P_{0}( \phi)$ and measurement events are independent, so $P( A_{1},B_{1} , A_{2},B_{2}|\phi )  =  P( A_{1},B_{1} |\phi )   P( A_{2},B_{2} |\phi ) $.

\section{Linear optics and parametric down-conversion} 

Now we translate our previous arguments into an optical context using an isomorphism between spins and a pair of harmonic oscillators (Schwinger representation). As such, one application that suggests itself is in measuring birefringence of optic fibers to high precision by comparing a test sample with a calibrated reference. The PDC process produces entangled photons by interaction of a pump laser field with a non-linear birefringent crystal. The output is shared among four optic modes, labeled $\{a_h,a_v, b_h,b_v\}$ for spatial directions $a,b$ and horizontal/vertical polarizations $h,v$. (We use the same notation to denote the associated bosonic annihilation operators.)  Applying the Schwinger representation to spatial mode $a$: $\hat{J}_{+a}\!=\!\hat{a}_{h}^{\dagger }\hat{a}_{v}$, $\hat{J}_{-a}\!=\!\hat{a}_{h}\hat{a}_{v}^{\dagger }$, and $2 \hat{J}_{z}\!=\! ({a}_h^{\dagger}{a}_h\!-\!{a}_v^{\dagger}{a}_v)$ (the number difference).  Also $\vec{J}_{a}^{2}\!=\!(\hat{n}_a/2) (\hat{n}_a/2\!+\!1)$ where $\hat{n}_a \!=\!   ({a}_h^{\dagger}{a}_h\!+\!{a}_v^{\dagger}{a}_v)$.  Spin quantum numbers map onto photon numbers as $2 j_a \!=\! (n_{ha}\!+\!n_{va}) $ and $2 m_a \!=\! ( n_{ha}\!-\! n_{va})$.  We can now identify each of the elements of our idealized parameter-estimation protocol with a realistic optical source, as explained in Fig.~\ref{Apparatus}.  Measurement data can be grouped by photon counting according to values of $j_{a,b}$, equivalent to post-selection onto the space of a particular $|\psi^{(j)}_{0}\rangle$. Optimal correlation measurements $ \hat{J}_{za} \otimes \hat{J}_{zb}$ are reconstructed in each $j_{a,b}$-labeled subspace, also from photon counting because $2 \hat{J}_{za}\!=\! \hat{n}_{ha}\!-\!\hat{n}_{va}$. We acknowledge here that efficient photon counting is generally a non-trivial task \cite{photoncounting}.

\section{Losses}

\begin{figure}[b]\centering
\includegraphics[width=3.2in]{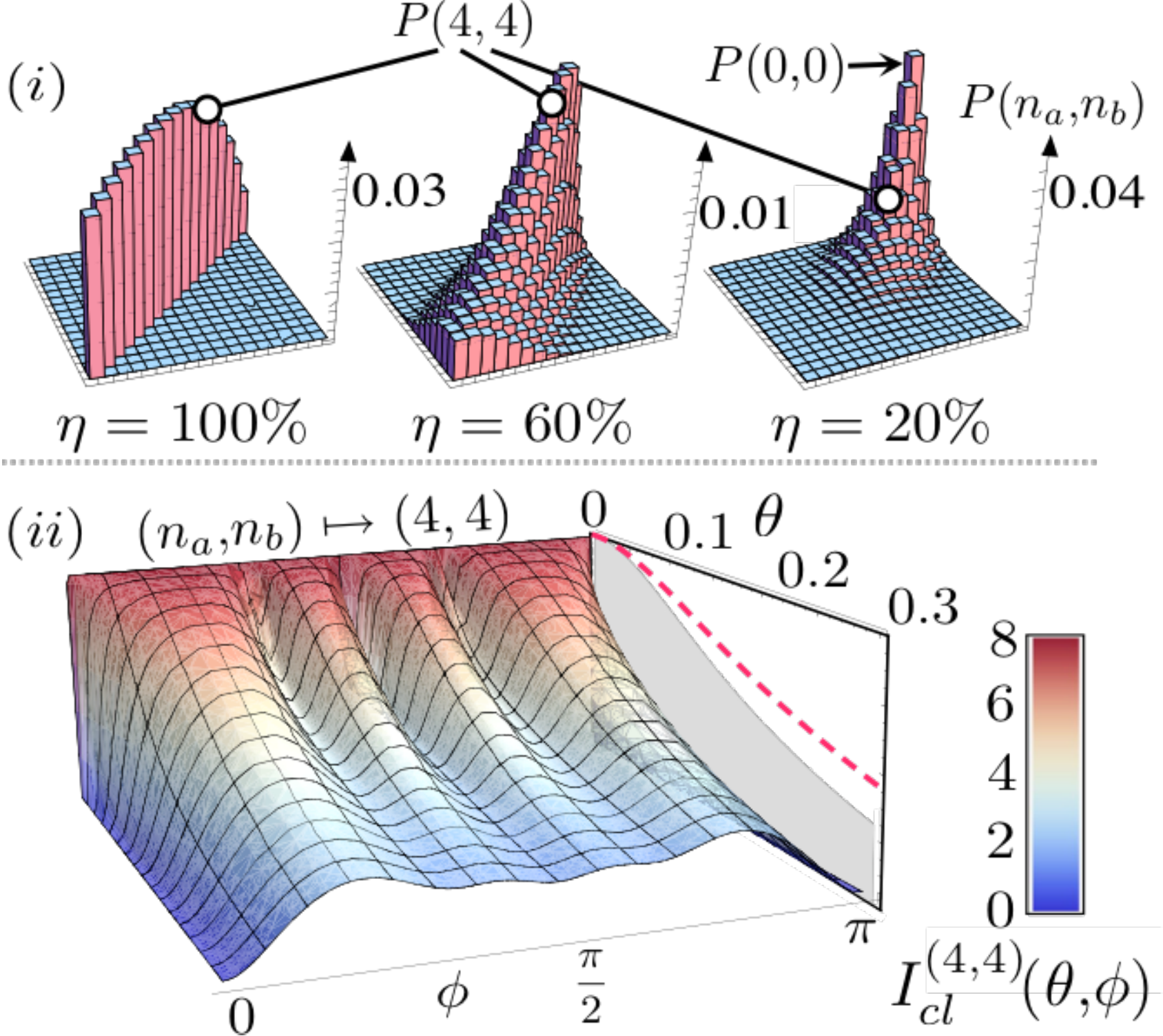}
\caption{\label{LossPlots}$ (i)$  Under photon loss the  PDC state becomes $\sum P(n_a, \! n_b) \rho^{(j_a,j_b)}$, where $j_a\!=\!n_a/2,j_b\!=\!n_b/2$, $ \rho^{(j_a,j_b)}$ is given in Eq.~(\ref{eqn:lossy_state}), and $P(n_a, \! n_b)$ is a weighting factor for the corresponding subspace \cite{Durkin04}.  The three plots show  $P(n_a,\! n_b)$ with interaction $\tau \sim 1.83$ and $n_{a,b} \leq 16$.  As transmission $\eta$ is reduced, the weightings of $j_a \neq j_b$ spaces increase and $n_{a,b} \gg 1$ become less probable. In $(ii)$ loss in transmission and detection affects the estimation of $\phi$ in the $(n_a, \! n_b)\! \mapsto \! (4,4)$ subspace.  The contribution to $\langle I_{cl}\rangle$ is $I_{cl}^{(4,4)}P(4,4)$.  For no losses ($\eta\!=\!1$), $\rho^{(2,2)}$ is the singlet state $\ket{\psi_0^{(2)}}  \bra{\psi_0^{(2)}} $ and $I_{cl}^{(4,4)}\!=\!8$, contributing
to the quadratic precision scaling derived for the lossless case.  As losses increase, the precision degrades and  $I_{cl}^{(4,4)}$ quickly becomes a function of both $\phi$ and the decoherence parameter $\theta\!=\! (1\!-\!\eta) \tanh \tau$.  Blue shading indicates precision below the upper limit for four uncorrelated photons in a classical lossless two-mode interferometer, i.e. $I_{cl}\leq 4$.  For $\phi \approx 0.5$ supra-classical precision (red shading) is possible for decoherence $\theta \lesssim 0.2$. This corresponds to loss $ \lesssim  26\%$ for $\tau = 1$, an interaction value achieved in real experiments \cite{Eisenberg04} . The maximum  $I_{cl}^{(4,4)}$ for each $\theta$ value with photon counting  measurements is projected on to the back wall as a grey silhouette -- only slightly inferior to  $I_{qu}^{(4,4)}$, the best possible for any detection scheme (red dashed line).}
\end{figure}

A realistic analysis must incorporate relevant decoherence; for optic processes incorporating photon counting the important mechanism is that of photon loss, in transmission and detection \cite{Durkin04}. Both loss types are effectively modeled by placing partial transmission ($\eta < 1$) beam-splitters in the four optic modes in front of perfect detectors, splitting incoming photons into two output modes:  the mode transformation is $\hat{a} \mapsto \sqrt{\eta} \hat{a}\!+\!\sqrt{1\!-\!\eta} \hat{e}$ where $\hat{e}$ is an annihilation operator for the ancillary loss mode. The photons syphoned out of the transmission mode in this way are then traced over. Referring either to modes $a$ or to modes $b$, if losses are polarization-insensitive $(\eta_{h}\!=\!\eta_{v})$, the loss channel ${\cal L}^h_\eta \otimes {\cal L}^v_\eta  $ commutes with any $U \in SU(2)$ on the same spatial path: $\mathcal{L}^{h}_{\eta} \otimes \mathcal{L}^{v}_{\eta} [U \rho U^{\dagger}] \!=\! U ( \mathcal{L}^{h}_{\eta} \otimes \mathcal{L}^{v}_{\eta} [ \rho] ) U^{\dagger}$ for any $\rho$.  This has two important consequences. Firstly, the non-unitary decoherence due to loss and the unitary $\phi$ rotation in mode $b$ may be treated independently with impunity. Secondly, after losses each component of the PDC state, in  a subspace labeled by $(j_a,j_b)$, retains its symmetry under transformations $U^{(j_a)}_a(g) \otimes U^{(j_b)}_b(g)$.  This implies a simple, block-diagonal structure for the mixed lossy state:
\begin{equation} \label{eqn:lossy_state}
\rho^{(j_a, j_b)}=\sum_{J=|j_a-j_b|}^{j_a+j_b}
\mu^{(j_a, j_b)}_J \Omega^{(j_a, j_b)}_J.
\end{equation}
The $\mu^{(j_a, j_b)}_J \in [0,1]$ are weighting factors \cite{Durkin04} and $\Omega^{(j_a, j_b)}_J$ are density operators proportional to identities in each $(2J+1)$-dimensional orthogonal subspace labeled by total spin $J$.  As symmetry is preserved under loss, the state of Eq.~(\ref{eqn:lossy_state}) retains a suitability for \emph{relative} measurements between spatially separated observers $a$ and $b$. We stress that for imperfect transmission $\eta <1$ there is non-zero occupancy probability for spaces labelled $(j_a,j_b)$ where $j_a \neq j_b$. See Fig.~\ref{LossPlots}(i). Measurements with $n_a \neq n_b$ need not be discarded, they also contribute to overall precision.

\section{Combining Subspaces and Ultimate Precision} 

By post-selecting onto specific photon numbers (after losses) and filtering the data sets we can focus on a particular $(j_a,j_b)$ subspace and analyze its contribution to overall precision. Fortunately, both $I_{qu}$ and $I_{cl}$ are additive, so total Fisher Information per measurement is the average of the contributions in each subspace weighted by the probabilities of post-selecting each subspace.  For the lossless case with $\hat{N}\!=\!\hat{n}_a\!+\!\hat{n}_b$ we know $ I_{qu}\!=\!4j(j\!+\!1)/3$ (in a space with $n_{a,b}\!=\!2j$ photons) and the ensemble result is $\langle I_{qu} \rangle\!=\!  \langle \hat{N}^2 \rangle /12\!+\! \langle \hat{N} \rangle /3$. There are associated gains in precision as $\langle \hat{N} \rangle\!=\! 4 \eta \sinh^2 \tau$ increases with $\tau$ in the low loss regime. For more significant loss there is a precision trade-off as larger $\tau$ are also  associated with greater values of decoherence $\theta\!=\! (1\!-\!\eta) \tanh \tau$ within each subspace. This may be viewed as higher--photon-number spaces being seeded initially, then as photons are lost these populations make an incoherent contribution to the those of lower photon spaces, `feeding' them from above with mixed state components (bad for precision).  We illustrate the effect of decreasing transmission  $\eta$  on both subspace weightings in Fig.~\ref{LossPlots}(i), and precision within a subspace, in Fig.~\ref{LossPlots}(ii) (explicitly the  $j_a\!=\!j_b\!=\!2$ subspace).

\begin{figure}[b]
\centering
\includegraphics[width=3.2in]{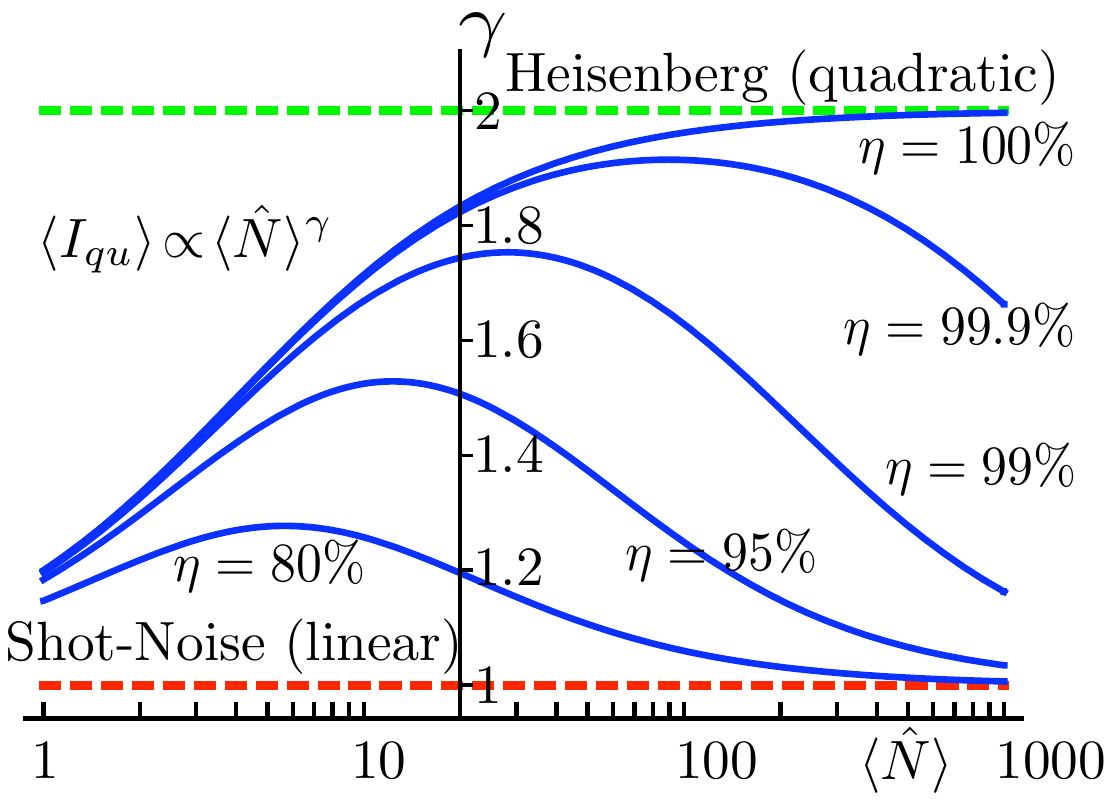}
\caption{\label{fluxpowerfig} The scaling of $\langle I_{qu} \rangle$ with the detected flux $\langle N \rangle$ is examined in the plot, for various values of the transmission parameter $\eta$.  A logarithmic scale is used for
$\langle N \rangle$ on the horizontal axis.  Scaling parameter $\gamma$ is defined by $\gamma\!=\!d  \log(\langle I_{qu}\rangle)/ d \log(\langle N\rangle) $, and locally $\langle I_{qu} \rangle \propto\langle N\rangle^\gamma$.  We see that for our protocol $\gamma$ varies between $1$ (shot-noise scaling) and $2$ (Heisenberg scaling), i.e. our scheme always scales better than the best classical interferometer using uncorrelated particles.  For moderate losses $\eta\approx 0.8$ the optimal flux is between $1$ and $10$ photons, well within the capabilities of some current number-resolving photodetection technologies \cite{photoncounting}.}
\end{figure}

Given a general mixed state $\rho \!=\! \sum_p \rho_p | p \rangle \langle p | $ Fisher Information for unitary evolution under $\hat{H} \mapsto \hat{J}_{yb}$ is
\begin{equation}
\label{eqn:qfiGeneral}
\langle I_{qu} \rangle \!=\! 2 \sum_{q,r} \frac{(\rho_q \!-\! \rho_r)^2}{\rho_q \!+\! \rho_r} \left| \langle q | \hat{J}_{yb}| r \rangle \right|^2 \; .
\end{equation}
By observing that the PDC is Gaussian, and that loss channel and interferometer components act as Gaussian operations, this functional can be evaluated directly using phase-space methods \cite{GaussianMethods}. Specifically, we employ the fact squeezed light subject to incoherent photon loss is formally equivalent to a squeezing operator (with a modified parameter) applied to a
thermal state.  This allows for the a full identification of the spectrum for the lossy PDC state, and for Eq.~(\ref{eqn:qfiGeneral}) to be evaluated directly by evaluating nonzero contributions to the sum (Appendix B).  We arrive at an expression for the complete Fisher information in terms of transmission $\eta$ and detection flux $\langle \hat{N} \rangle = 4 \eta \sinh^2 \tau$:
\begin{equation}\label{Hugo}
\frac{\langle \hat{N}^2 \rangle \!+\!  4 \langle \hat{N} \rangle}{12} \geq  \langle I_{qu} \rangle \!=\!  \frac{\langle \hat{N} \rangle \left( 4 \eta \!+\! \langle \hat{N} \rangle  \right)}{8 \!+\! 4 (1\!-\!\eta) \langle \hat{N} \rangle}  > \frac{\langle \hat{N} \rangle^2 }{8 \!+\! 4 \langle \hat{N} \rangle}
\end{equation}
where the upper and lower limits correspond to $\eta = 1,0$ respectively. Therefore, even in the worst case, losses approaching $100 \%$, \emph{any} non-zero flux gives $\langle I_{qu} \rangle \sim  \langle \hat{N} \rangle/4$ and shot noise limited precision scaling $ \delta \phi \sim \langle \hat{N} \rangle^{-1/2} $ (in stark contrast to the exponential deterioration in performance of other schemes \cite{Chen07}). As losses increase, flux $ \langle \hat{N} \rangle$ can be maintained by turning up the interaction $\tau$, but the Fisher information will inevitably deteriorate because of its separate dependence on $\eta$ and  $ \langle \hat{N} \rangle$ above. It is not known whether optimal measurements in the lossy case can be independent of the true parameter value $\phi$, certainly photon correlation measurements $\hat{J}_{za} \otimes \hat{J}_{zb}$ are no longer optimal, see Fig.~\ref{LossPlots}$(ii)$. From Eq.~\eqref{Hugo} it is seen that our scheme performs better than a sample illuminated with coherent light of the same detected flux, for which  $\langle I_{qu} \rangle\!=\! \langle \hat{N} \rangle/2$,
when $ \eta > 1/2+1/(2\!+\!\langle \hat{N} \rangle)$.  Transmission must be better than $50 \%$ in the high flux limit to obtain a quantum advantage with PDC. Fig.~ \ref{fluxpowerfig} shows the scaling of $\langle I_{qu} \rangle$ with detected flux $\langle \hat{N} \rangle$ when fitted locally to a power law: $\langle I_{qu} \rangle \propto \langle \hat{N} \rangle^{\gamma}$.

\section{Summary and Outlook}

We have presented a parameter-estimation protocol with several strengths.  In a non-dissipative environment,  Heisenberg-scaling is achieved with simple fixed measurements.  Our scheme can be implemented using PDC and linear optics, and under severe dissipation (approaching $100 \%$) it is capable of precision scaling that approaches the shot-noise limit asymptotically \emph{from above}.  There exist a variety applications, such as fiber calibration, the phase microscopy of fragile biological specimens, and optical gyroscopes for GPS-free navigation. We will, in a forthcoming work, extend this approach beyond a single phase estimation to all of $SU(2)$, parameterized by three Euler angles, and utilizing multiple spin correlations.  The increasing asymmetry of the state with the magnitude of one-sided rotations has a quantifiable utility in reference frame alignment \cite{REoFr}.

\section{Acknowledgements}
The authors thank Antia Lamas-Linares, Terry Rudolph, Vadim Smelyanskiy and Alessio Serafini for helpful discussions. H.C. acknowledges support for this work by the National Research Foundation and Ministry of Education, Singapore.  G.A.D. performed this work under contract with Mission Critical Technologies, Inc.

\appendix

\section{Derivation of $I_{cl}$ for pure-state probes and no dissipation}

 Referring to the parameter-estimation paradigm described in the main text, we set $\hat{H} \mapsto \hat{J}_{yb}$ and $t \mapsto \phi$, and $\hat{M} \mapsto \hat{J}_{za} \otimes \hat{J}_{zb}$.  Under the unitary dynamics, the $j$-singlet probe state evolves to $|\psi_\phi^{(j)} \rangle\!=\!\exp \left( -i \phi \hat{J}_{yb} \right) | \psi _{0}^{(j)} \rangle$.  There are $(2j\!+\!1)^2$ outcomes, and the probability distribution for these is of the form $P_{AB}\left( \phi \right) =\left\vert \left\langle j,A\right\vert _{z,a}\left\langle j,B\right\vert _{z,b}\left\vert \psi_{\phi} \right\rangle \right\vert ^{2}\!=\!d_{B,-A}^{\left( j\right) }\left( \phi \right) ^{2}/\left( 2j+1\right)$.

The classical Fisher information is defined to be, $I_{cl}\left(\phi\right) \!=\!\sum_{A,B}P_{A,B}\left[ d \ln \left( P_{A,B}\right)/d\phi  \right] ^{2}$.  The derivatives $dP_{AB}/d\phi$ are given by $\left[d^{\left( j\right)}_{B,-A}\left( \phi \right)/\left( 2j+1\right)\right]\times$ $\left[ N^{(j)}_{-}\left(-A\right) d_{B,-A-1}^{\left( j\right) }\left( \phi \right) \!-\!N^{(j)}_{+}\left( -A\right) d_{B,-A+1}^{\left( j\right) }\left( \phi \right) \right]$.  Summing over all detection outcomes gives $I_{cl}\left( \phi \right) =4j\left( j\!+\!1\right)/ 3$, and this is independent of the value of $\phi$ and equal to $I_{qu}$.

\section{$\langle I_{qu} \rangle $ for the PDC state in the case of equal loss rates for every mode}

Referring to the parameter-estimation paradigm described in the main text, we set $\hat{H} \mapsto \hat{J}_{yb}$ and $t \mapsto \phi$.  Now an optimal measurement is assumed, but is not made explicit, and we derive the best precision possible in principle.  Our probe state is assumed to be the PDC state corresponding to effective interaction time $\tau$.  Photon loss arising from transmission and detection is assumed to act independently on all four optic modes, and the same transmissivity parameter $\eta$ is assumed in each case.

The calculation of $\langle I_{qu} \rangle$ is presented in four stages: (i) the lossless PDC state is written in terms of one-mode squeezing operations; (ii) the action of the loss super-operator on the squeezed vacuum is explained; (iii) eigenvalues and eigenkets for the lossy PDC state are identified; (iv) $\langle I_{qu} \rangle$ is evaluated in terms of $\tau$, $\eta$.

{\it (i) The lossless PDC state represented as a Gaussian state:} The PDC state is of the form,
\be
|\psi _{\rm pdc}\rangle \propto \sum_{n=0}^{\infty }\sqrt{n+1}\lambda ^{n}e^{in\varphi }|\psi _{0}^{\left( n/2\right) }\rangle, \nonumber
\ee
where the phase factor $\varphi$ is the phase of the pump field, $\lambda\!=\!\tanh \left( \tau \right)$ corresponds to the strength of the non-linear interaction, and the normalized photonic $n/2$-singlet state is $|\psi _{0}^{\left( n/2\right) }\rangle\propto$ $\sum_{m=0}^{n}\left( -1\right) ^{m}|n-m,m,m,n-m\rangle _{a_{h}a_{v}b_{h}b_{v}}$.

Next, we rewrite the PDC state in terms of Gaussian operations acting on the vacuum, and specifically in terms of squeezing processes acting in single modes:
\bea
|\psi_{\rm pdc} \rangle
&\!=\!&\hat{U}_{\text{global}}(\varphi )\hat{U}_{a_{h}b_{v}}^{\text{bs}}\hat{U}_{a_{v}b_{h}}^{\text{bs}} \nonumber \\
&& \, \circ\, \hat{S}_{a_{h}}\left( \tau \right) \hat{S}_{a_{v}}\left( -\tau \right)  \hat{S}_{b_{h}}\left( \tau \right) |0\rangle \hat{S}_{b_{v}}\left( -\tau \right) \ket{0}, \nonumber
\eea
where
$\hat{U}_{ab}^{\text{bs}}\!=\!\exp \left[ \left( \pi /4\right) \left( \hat{a}^{\dag }\hat{b}-\hat{a}\hat{b}^{\dag }\right) \right] $
denotes a 50:50 beam-splitter operation, $\hat{U}_{\text{global}}(\varphi )\!=\!\exp (i\hat{N}\varphi /2)$ accounts for the global phase factor, and the squeezing operations are defined by
$\hat{S}_{a_{h}}\left( \tau \right)\!=\!\exp \left[ \!-\!\left( \tau /2\right) \left( \hat{a}_{h}^{\dag 2}\!-\!\hat{a}_{h}^{2}\right) \right] $,
$\hat{S}_{a_{v}}\left( -\tau \right)\!=\!$ $\exp \left[ \left( \tau /2\right) \left( \hat{a}_{v}^{\dag 2}\!-\!\hat{a}_{v}^{2}\right) \right] $,
and similarly for modes $b$.

{\it (ii) The loss channel as a Gaussian operation and its action on the squeezed vacuum state:}
The loss super-operator ${\cal L}^c_{\eta}$ for mode $c \in \{a_h,a_v, b_h,b_v\}$ acts as
${\cal L}^c_{\eta}[\rho]=\sum
\limits_{n=0}^{\infty} L_n^c \rho L_n^{c\dagger},$ with $L_n^c=(1-\eta)^{n/2} c^n
\eta^{c^{\dagger}c/2}/\sqrt{n!} $, where the loss in mode $c$ is $(1-\eta) \in [0,1]$.  We note that ${\cal L}^c_{\eta}$ is a Gaussian operation,
and therefore that the action of ${\cal L}^c_{\eta}$ on the squeezed vacuum state for one mode can readily be computed using standard continuous-variable methods.  We make the following formal identification:
\be
{\cal L}^c_{\eta}[\hat{S}^{\left( 1\right) }\left( \pm \tau \right) |0\rangle ]=\hat{S}\left( \pm \tau _{\rm eff}\right) \hat{\rho}_{\rm th}(\bar{N})\hat{S} ^\dag\left( \pm \tau _{\rm eff}\right) , \nonumber
\ee
where $\hat{\rho}_{\rm{th}}$ denotes the thermal mixed state $\hat{\rho}_{\rm{th}} \propto \sum_{n=0}^{\infty }\chi ^{n}|n\rangle \langle n|$, in the Fock basis, and $\chi =\bar{N}/\left( 1+\bar{N}\right)$.
The thermal intensity parameter $\bar{N}$ and the effective squeezing parameter $\tau_{\rm eff}$ are related to the PDC and transmission parameters as follows:
\bea
\tau _{\rm eff}&=&\frac{1}{4}\ln \left( \frac{P}{M}\right), \nonumber \\
\bar{N}&=&\frac{-1+\sqrt{PM}}{2}, \nonumber
\eea
$P= \eta e^{2\tau }+1-\eta $, and $M=\eta e^{-2\tau }+1-\eta$.

\newpage
{\it (iii) Eigenvalues and eigenvectors of the lossy PDC state:}
Since the photon losses that arise due to transmission and detection are assumed to be polarization insensitive, they commute with the interferometric stages of our protocol, and can therefore be assumed to act exclusively at the PDC source.  Furthermore, because the same value for the transmission parameter $\eta$ is assumed for all four modes, the PDC state can be written as follows, after accounting for all losses:
\be
\hat{\rho}_{\rm PDC}^{\rm lossy}\!=\!\hat{U}_{\rm eff}\left [\hat{\rho}_{\rm th}(\bar{N})^{\otimes 4}\right]\hat{U}_{\rm eff}^\dag, \nonumber
\ee
where,
\bea
\hat{U}_{\rm eff}&=&\hat{U}_{\rm global}(\varphi )\hat{U}_{a_{h}b_{v}}^{\rm bs}\hat{U}_{a_{v}b_{h}}^{\rm bs} \nonumber \\
&& \circ
\hat{S}_{a_{h}}\left( \tau _{\rm eff}\right) \hat{S}_{a_{v}}\left( -\tau _{\rm eff}\right) \hat{S}_{b_{h}}\left( \tau _{\rm eff}\right) \hat{S}_{b_{v}}\left( -\tau _{\rm eff}\right). \nonumber
\eea
Definitions for $\hat{U}^{\rm bs}$, $\hat{U}_{\rm global}$ and $\hat{S}$ are as in subsection (i).  Parameters $\tau_{\rm eff}$ and $\bar{N}$ are as in subsection (ii).

The eigenkets of $\hat{\rho}_{\rm PDC}^{\rm lossy}$  are therefore given by $\hat{U}_{\rm eff} \ket{A_h,A_v,B_h,B_v}$, for every product of Fock states in four modes: $\ket{A_h,A_v,B_h,B_v}_{a_h a_v b_h b_v}$.  The corresponding eigenvalues are of the form
$\chi ^{A_{h}+A_{v}+B_{h}+B_{v}}/\left( 1+\bar{N}\right) ^{4}$.

{\it (iv) Evaluating $\langle I_{qu} \rangle$ for the lossy PDC state:} For a mixed probe state $\hat{\rho}\left( \phi \right)$, subject to a unitary evolution $\exp \left( -i\phi \hat{H}\right) $, the quantum Fisher information is given by $\langle I_{qu} \rangle \!=\!{\rm tr}\left\{ \left( d\hat{\rho}/d\phi \right) R_{\hat{\rho}}^{-1}\left[ d\hat{\rho}/d\phi \right] \right\}$, where $R_{\hat{\rho}}^{-1}\left[ d\hat{\rho}/d\phi \right]$ denotes the symmetric logarithmic derivative of $\hat{\rho}$. The value of $\langle I_{qu} \rangle $ is independent of $\phi$, and we set $\phi=0$.  By diagonalizing the probe state, $\hat{\rho} \!=\! \sum_p \rho_p | p \rangle \langle p | $, the quantum fisher information can be evaluated using the expression,
\begin{equation}
I_{qu}  = 2 \sum_{q,r} \frac{(\rho_q - \rho_r)^2}{\rho_q + \rho_r} \left| \langle q | \hat{H}| r \rangle \right|^2. \nonumber
\end{equation}
In what follows
$\hat{H}\mapsto\left( \hat{b}_{v}\hat{b}_{h}^{\dagger }-\hat{b}_{v}^{\dagger }\hat{b}_{h}\right)/2i $,
corresponding to the Hamiltonian for a $y$-axis rotation in modes $b$, while the eigenkets and eigenvalues are those, given in subsection (iii), for the PDC state after the loss super-operator has acted on all four modes.

To evaluate $\langle I_{qu} \rangle$, we treat it in stages.  There are eight summing indices, $A_h$,$A_v$,$B_h$,$B_v$ and
$A^\prime_h$,$A^\prime_v$,$B^\prime_h$,$B^\prime_v$, corresponding to possible photon numbers in the four modes.
Next,
\bea
\frac{(\rho_q - \rho_r)^2}{\rho_q + \rho_r}
&\mapsto& \frac{1}{\left( 1+\bar{N}\right) ^{4}} \nonumber \\
&\times &
\frac{\left( \chi ^{A_{h}+A_{v}+B_{h}+B_{v}}\!-\!\chi ^{A_{h}^{\prime }+A_{v}^{\prime }+B_{h}^{\prime }+B_{v}^{\prime }}\right) ^{2}}{ \chi ^{A_{h}+A_{v}+B_{h}+B_{v}}+\chi ^{A_{h}^{\prime }\!+\!A_{v}^{\prime }+B_{h}^{\prime }+B_{v}^{\prime }} }. \nonumber
\eea
By acting $U_{\rm eff}$ on the Hamiltonian, the final component of the sum is found to be given by,
\bea
\left| \langle q | \hat{H}| r \rangle \right|^2 &\mapsto&
\frac{1}{16}\sinh ^{2}\left( 2\tau _{\rm eff}\right) \nonumber \\
&\times&
\left\vert \langle A_{h}^{\prime }A_{v}^{\prime }B_{h}^{\prime }B_{v}^{\prime }
|\hat{T}|A_{h}A_{v}B_{h}B_{v}\rangle \right\vert ^{2}, \nonumber
\eea
where $\hat{T}\!=\! -\!\hat{a}_{h}\hat{a}_{v}\!+\!\hat{b}_{v}\hat{b}_{h}\!+\!\hat{a}_{h}^{\dag }\hat{a}_{v}^{\dag }\!-\!\hat{b}_{v}^{\dag }\hat{b}_{h}^{\dag }$, and terms which do not contribute to the sum have been dropped.  For a given value of the summing indices, at most one
component of $\hat{T}$ can generate a nonzero contribution.  Hence,
$\left\vert \langle A_{h}^{\prime }A_{v}^{\prime }B_{h}^{\prime }B_{v}^{\prime }
|\hat{T}|A_{h}A_{v}B_{h}B_{v}\rangle \right\vert ^{2}
=
\left\vert \langle A_{h}^{\prime }A_{v}^{\prime }B_{h}^{\prime }B_{v}^{\prime }|
\hat{a}_{h}\hat{a}_{v}|A_{h}A_{v}B_{h}B_{v}\rangle \right\vert ^{2} +
$ three similar terms.  Furthermore, each of these four terms provide an equal total contribution to the sum.

Now $\langle I_{qu} \rangle$ can be evaluated directly:
\bea
\langle I_{qu} \rangle &=&\frac{\sinh ^{2}\left( 2\tau _{\text{eff}}\right) \left( 2\bar{N}+1\right) ^{2}}{2\left( 1+2\bar{N}+2\bar{N}^{2}\right) } \nonumber \\
&=& \frac{\left[ \eta \cosh ^{2}\left( \tau \right) \right] \left[ 2\eta \sinh ^{2}\left( \tau \right) \right] }{1+\left( 1-\eta \right) \left[ 2\eta \sinh ^{2}\left( \tau \right) \right] }\,. \nonumber
\eea
Finally, it can be recast in terms of the total detected flux $\langle \hat{N} \rangle = 4 \eta \sinh^2 \tau$,
\be
\langle I_{qu} \rangle  \!=\!  \frac{\langle \hat{N} \rangle \left( 4 \eta \!+\! \langle \hat{N} \rangle  \right)}{8 \!+\! 4 (1\!-\!\eta) \langle \hat{N} \rangle}\,. \nonumber
\ee
Using this result, the scaling of $\langle I_{qu}\rangle $ as a function of the detected flux and the transmission parameter is examined in Fig.\ref{fluxpowerfig}.


\begin{thebibliography}{99}
\bibitem{notation} We use a standard notation for angular momentum operators $\hat{J}_i$ for which $\hat{J}_i | j,m \rangle_i = m | j,m \rangle_i$.  Operator $\vec{J}^2\! =\! \sum_{i \in \{x,y,z\}}\hat{J}_i^2$ gives  $\vec{J}^2 | j,m \rangle_i = j(j+1)| j,m \rangle_i$ and labels subspaces invariant under the action of $SU(2)$.   Ladder operators are defined by $\hat{J}_\pm\!\!=\!\!\hat{J}_x \!\pm\! i \hat{J}_y$, and $\hat{J}_{\pm }\left\vert j,m\right\rangle_z\!\!\!=\!\!\!N^{(j)}_{\pm }\left( m\right) \left\vert j,m\!\pm\! 1\right\rangle_z$ where $N^{(j)}_{\pm }\left( m\right)\!=\!\sqrt{\left( j\!\mp\! m\right) \left( j\!\pm\! m\!+\!1\right) }$.   Wigner's rotation matrix is $d_{m^{\prime },m}^{\left( j\right) }\left( \phi \right)\!=\!\left\langle jm^{\prime }\right\vert_z \exp \left( -i\hat{J}_{y}\phi \right) \left\vert jm\right\rangle_z$.
\end{thebibliography}
\end{document}